\documentclass[a4paper,twoside,10pt]{letter}
\usepackage{amsmath,amssymb}
\usepackage{graphicx,saj,multicol,subeqnarray}
\usepackage{soul,astro}

\def\arcmin{\hbox{$^\prime$}}
\def\arcsec{\hbox{$^{\prime\prime}$}}

\def\papertype{Original Scientific Paper}

\def\udc{...}
\begin{document}
\baselineskip=3.1truemm
\columnsep=.5truecm
\newenvironment{lefteqnarray}{\arraycolsep=0pt\begin{eqnarray}}
{\end{eqnarray}\protect\aftergroup\ignorespaces}
\newenvironment{lefteqnarray*}{\arraycolsep=0pt\begin{eqnarray*}}
{\end{eqnarray*}\protect\aftergroup\ignorespaces}
\newenvironment{leftsubeqnarray}{\arraycolsep=0pt\begin{subeqnarray}}
{\end{subeqnarray}\protect\aftergroup\ignorespaces}
\newcommand{\SNR}{\mbox{{G1.9+0.3}}}
\def\HI{\hbox{H\,{\sc i}}}
\def\HII{\hbox{H\,{\sc ii}}}
\def\p0{\phantom{0}}
\def\degr{\hbox{$^\circ$}}
\def\arcmin{\hbox{$^\prime$}}
\def\arcsec{\hbox{$^{\prime\prime}$}}
\newcommand{\D}{$^\circ$}
\newcommand{\farcm}{\mbox{\ensuremath{.\mkern-4mu^\prime}}}
\newcommand{\farcs}{\mbox{\ensuremath{.\!\!^{\prime\prime}}}}
\newcommand{\fdg}{\mbox{\ensuremath{.\!\!^\circ}}}


\markboth{\eightrm Radio-continuum Emission from the Young Galactic SNR \SNR} {\eightrm A.~Y.~DE~HORTA, et. al.}

{\ }

\publ

\type

{\ }


\title{Radio-continuum Emission from the Young Galactic Supernova Remnant \SNR}


\authors{A.~Y.~De~Horta$^{1}$, M.~D.~Filipovi\'c$^{1}$, E.~J.~Crawford$^{1}$, F.~H.~Stootman$^{1}$, T.~G.~Pannuti$^{2}$,}
\authors{L.~M.~Bozzetto$^{1}$, J.~D.~Collier$^{1}$,  E.~R.~Sommer$^{1}$ and A.~R.~Kosakowski$^{2}$}

\vskip3mm


\address{$^{1}$University of Western Sydney, Locked Bag 1797, Penrith South, DC, NSW 1797, Australia }

\Email{a.dehorta}{uws.edu.au, m.filipovic@uws.edu.au, e.crawford@uws.edu.au}

\address{$^{2}$Space Science Center, Department of Earth and Space Sciences, Morehead State University, 235 Martindale Drive, Morehead, KY 40351 USA}

\Email{t.pannuti}{moreheadstate.edu}


\dates{June 5, 2014}{June 18, 2014}


\summary{We present an analysis of a new Australia Telescope Compact Array (ATCA) radio-continuum observation of supernova remnant (SNR) \SNR, which at an age of \mbox{$\sim$181$\pm$25} years is the youngest known in the Galaxy. We analysed all available radio-continuum observations at 6-cm from the ATCA and the Very Large Array. Using this data we estimate an expansion rate for \SNR\ of 0.563\%$\pm$0.078\% per year between 1984 and 2009. We note that in the 1980's \SNR\ expanded somewhat slower (0.484\% per year) than more recently (0.641\% per year). We estimate that the average spectral index between 20-cm and 6-cm, across the entire SNR is \mbox{$\alpha$=--0.72$\pm$0.26} which is typical for younger SNRs. At 6-cm, we detect an average of 6\% fractionally  polarised radio emission with a peak of 17\%$\pm$3\%. The polarised emission follows the contours of the strongest of \mbox{X-ray} emission. Using the new equipartition formula we estimate a magnetic field strength of B$\approx$273~$\mu\mbox{G}$, which to date, is one of the highest magnetic field strength found for any SNR and consistent with \SNR\ being a very young remnant.
}



\begin{multicols}{2}


\section{1. INTRODUCTION}

It is widely accepted that current catalogues have a distinct deficit of young Galactic supernova remnants (SNRs), that is, SNRs $<$2000 years old, with only $\sim$10 confirmed out of a predicted $\sim$50 (van den Bergh \& Tammann 1991; Cappellaro 2003). Of these confirmed SNRs, \SNR\ is of particular interest as it is believed to be the youngest in the Milky Way (MW) at $\sim150$ years old (Reynolds et al. 2008; Green et al. 2008; Reynolds et al. 2009; Carlton et al. 2011a,b).

Originally identified as a probable SNR by Green \& Gull (1984) at 4.9~GHz using the Karl G. Jansky Very Large Array (VLA), \SNR\ was described as a shell source with an approximate brightness slightly less than that of the Tycho and Kepler SNRs with a spectral index of $\alpha\sim-0.7$\footnote{Spectral index defined as $S\propto\nu^\alpha$}. Using the Molonglo Observatory Synthesis Telescope (MOST) Galactic Survey data, Gray (1994) confirmed the classification of \SNR\ as an SNR: the source was described as featuring a shell-like morphology in the radio with an estimated diameter of 1\farcm2. Later, LaRosa et al. (2000) produced a 90-cm image of \SNR\ made using observations from the VLA. They estimated the 20/90-cm spectral index of the SNR to be $\alpha=-0.93\pm0.25$ and the angular  diameter to be 1\farcm1. Nord et al. (2004) revisited the data collected by LaRosa et al. (2000) and -- through the application of superior data reduction techniques -- measured the diameter of \SNR\ to be $<$1\arcmin\ and placed it at a distance of $<$7.8~kpc. Green (2004) estimates the diameter of \SNR\ based on 1.49~GHz VLA observations made in 1985 to be 1\farcm2. Most recently, Roy \& Pal (2014) measured a \HI\ absorption distance using known anomalous velocity features near the Galactic Centre (GC) and found a lower limit on \SNR\ distance from Sun as 10~kpc, some 2~kpc further away from the GC. Therefore, multiple radio observations have confirmed that \SNR\ has the smallest angular diameter for a known Galactic SNR, indicative of its young age.

Green et al. (2008) re-observed \SNR\ at 4.86~GHz using the VLA after Reynolds et al. (2008) used 2007 {\it Chandra} images to show \SNR\ had expanded significantly since 1985 and its X-ray emission appeared  to be predominantly synchrotron in nature. By comparing these new VLA observations with the 1985 VLA observations made at 1.49~GHz, Green et al. (2008) determined that \SNR\ had expanded by $15\%\pm2\%$ over 23 years ($\sim$0.65$\%$ per year). Using the same VLA observations from 1985 and 1989, G\'omez \& Rodr\'iguez (2009) derived an expansion rate of $0.46\%\pm0.11\%$ and an age of $220\pm ^{70}_{45}$ years. By comparing 2007 and 2009 {\it Chandra} X-ray images and utilising a simple uniform-expansion model, Carlton et al. (2011a,b), find an expansion rate of 0.642\% $\pm$ 0.049\% yr$^{-1}$ and a flux increase of 1.7\% $\pm$ 1.0\%~yr$^{-1}$, ageing the remnant at 156 $\pm$ 11~yr assuming no deceleration. Murphy, Gaensler \& Chatterjee (2008) found that \SNR's flux density at 843~MHz increased by $1.22\pm ^{0.24}_{0.16}$ \% per year over the last two decades.

Borkowski et al. (2013) suggest that \SNR\ was likely a Type~Ia SNe with the shell of its remnant in free expansion with a velocity $\sim$18\,000~km~s$^{-1}$. The ejecta shows spatial asymmetry with prominent Fe-group elements in the northern rim. Also, we point out that Abramowski et al. (2014) report no  $\gamma$-ray signal from \SNR\ using observations from the H.E.S.S. (High Energy Stereoscopic System) Cherenkov telescope array. 

The presence of polarised emission and spatial spectral variations are identified by Farnes (2012), with flatter spectra identified in the NW and SE of the remnant.

In this paper, we present the results of our Australia Telescope Compact Array (ATCA) radio-continuum observations of \SNR\ made at 20, 13 \& 6~cm in 2009. A comprehensive expansion study is conducted by comparing the new 6-cm observations with a previously unpublished 6-cm ATCA radio-continuum observation, made in 1993, and three 6-cm VLA observations made in 2008, 1989 and 1984 respectively. Here, we also report, on the radio-continuum spectral energy distribution and polarisation properties of this young SNR.

\section{2. OBSERVATIONAL DATA}

\SNR\ was observed at 20, 13 \& 6~cm wavelengths on four days in 2009 (Project C1952). Two of those days being in January 2009 with the remainder in February 2009. The 20-cm and 13-cm observations taken on the 2nd and 3rd day, were carried out simultaneously as they make use of a common feed-horn. The 6-cm observations taken on the 1st and 4th day, used four different frequencies (two per day) to improve multi-frequency synthesis (MFS). For this purpose, the \textsc{miriad} (Sault \& Killeen 2008) task \mbox{\textsc{mfplan}} was used to select the most appropriate frequencies. Over the four days \SNR\ was observed with two separate antenna configurations for a total of 29 independent baselines covering a range of spacings from 31 to 6\,000~m. See Table~1 for complete observational details.

\end{multicols}

\noindent
\parbox{\columnwidth}{
{\bf Table 1.} 2009 ATCA observations of \SNR. \\
\vskip.25cm
\centerline{
\begin{tabular}{lcccc} 
 \hline
                        & Day 1        & Day 2        & Day 3        & Day 4\\
 \hline       
  Date                  & 03 Jan       & 04 Jan       & 06 Feb       & 07 Feb \\
  ATCA Array            & 6C           & 6C           & EW352        & EW352         \\
  Frequency 1           & 4.672~GHz    & 1.384~GHz    & 1.384~GHz    & 4.544~GHz\\
  Frequency 2           & 5.440~GHz    & 2.368~GHz    & 2.368~GHz    & 5.184~GHz\\
  Bandwidth             & 128~MHz      & 128~MHz      & 128~MHz      & 128~MHz\\
  Time on source        & 332~min & 458~min & 509~min & 1033~min \\
  Primary Calibrator    & J1934-638    & J1934-638    & J1934-638    & J1934-638 \\
  Secondary Calibrator  & J1741-312    & J1751-253    & J1751-253    & J1741-312\\
  \hline
\end{tabular}}} 
\vskip.5cm

\begin{multicols}{2}

\centerline{
\includegraphics[width=0.45\textwidth,angle=-90]{fig1}}
\label{fig1}
\figurecaption{1.}{ATCA 20-cm image of Galactic SNR \SNR. The blue ellipse in the lower left corner represents the synthesised beam of $10\farcs9 \times 5\farcs4$ at PA=--$0\fdg5$. Contours are drawn at $3\sigma$, $5\sigma$, $8\sigma$, $12\sigma$, $17\sigma$, $23\sigma$, $30\sigma$, $38\sigma$ $47\sigma$ \& $57\sigma$ ($\sigma=0.22$~mJy/beam) 
}

\centerline{
\includegraphics[width=0.45\textwidth,angle=-90]{fig2}}
\label{fig2}
\figurecaption{2.}{ATCA 13-cm image of Galactic SNR \SNR. The blue ellipse in the lower left corner represents the synthesised beam of $6\farcs1 \times 2\farcs9$ at \mbox{PA=--$0\fdg5$.} Contours are drawn at $3\sigma$, $5\sigma$, $8\sigma$, $12\sigma$, $17\sigma$, $23\sigma$, $30\sigma$, $38\sigma$ $47\sigma$ \& $57\sigma$ ($\sigma=0.32$~mJy/beam) 
}

In Figs.~1 through 3 we show our new ATCA 2009 images of \SNR\ at 20-cm, 13-cm and 6-cm respectively. All these images were formed using MFS with uniform weighting and were deconvolved using the \textsc{miriad} (Sault \& Killeen 2008) \textsc{clean} and \textsc{restor} tasks, with self calibration being applied to the 6-cm image only. We note that our corresponding ATCA  flux density measurements are significantly smaller ($\sim$50\%) than the VLA estimates of Green et al. (2008). We can attribute this large difference to missing short spacings and poorer {\it uv} coverage of the ATCA images.


\centerline{
\includegraphics[width=0.45\textwidth,angle=-90]{fig3}}
\label{fig3}
\figurecaption{3.}{ATCA 6-cm image of Galactic SNR \SNR. The blue ellipse in the lower left corner represents the synthesised beam of $2\farcs8 \times 1\farcs2$ at \mbox{PA=--$0\fdg5$.} Contours are drawn at $3\sigma$, $5\sigma$, $8\sigma$, $12\sigma$, $17\sigma$, $23\sigma$, $30\sigma$, $38\sigma$ $47\sigma$ \& $57\sigma$ ($\sigma=0.07$ mJy/beam) 
}


\section{3. RESULTS}

 \subsection{3.1 \SNR\ Expansion and Age}
 \label{sec:exp}

A simple way to determine the age (in years) of a young SNR is to compare two images of the SNR taken at different epochs and to measure the percentage~expansion the SNR has undergone over the time between the observations. Since we know that the SNR will have expanded by 100\% in the intervening period between the SN explosion and the later observation, we can apply the simple formula, $Age=100\%/ER$, where $ER$ is the percentage~expansion the SNR has undergone over the time between the observations (in years). 

Ideally, to most accurately determine the expansion rate of an SNR, one should compare images from similar observations i.e, at the same wavelength and having similar, if not identical \emph{uv}-coverage, resolution and $rms$ noise. 

In the case of \SNR\ the first observations were made in 1984 using the VLA, with the best image produced from the 6~cm data (Fig.~4, bottom left). With this in mind, ATCA and VLA archives 

\end{multicols}

\noindent
\parbox{\columnwidth}{
{\bf Table 2.} VLA observations of  \SNR. \\
\vskip.25cm
\centerline{
 \begin{tabular}{@{}lccc}
 \hline
   & 1984 obs. & 1989 obs. & 2008 obs.\\
 \hline       
  Date & 26 May 1984 & 23 June 1989 & 12 March 2008 \\
  VLA Array  & C & BC & C \\
  Frequency & 4.89~GHz & 4.89~GHz & 4.89~GHz \\
  Bandwidth & 50~MHz & 50~MHz & 50~MHz \\
  Time on source & $\sim$10~min & $\sim$10~min & $\sim$30~min \\
  Primary Calibrator  & 3C286 & 3C286 & 3C286 \\
  Secondary Calibrator  & J1832-105 & J1751-253 & J1751-253 \\
\hline        
\end{tabular}}} 
\vskip.5cm

\begin{multicols}{2}

\noindent were searched for all available radio-continuum data of \SNR\ made at this same wavelength --- 6-cm. 

For the expansion study carried out in this paper, one ATCA and three VLA observations made at 6-cm were found at different epochs from the original 1984 to 2008.

The archival ATCA observation were taken on the 10$^{th}$ June 1993 (Project C034; P.I.: A. Gray). This observation was made in the 6-cm band centred at 4672 and 5440~MHz with a bandwidth (BW) of 128~MHz and the telescope in 6A configuration\footnote{We combined these two frequencies into a single 1~GHz band, see image shown in Fig.~4 (middle left).}. Source 1934-638 was used for primary calibration and source 1748-253 was used for secondary calibration. The observations were done in so-called ``snap-shot'' mode, totalling $\sim$1~hour of integration spread equally over a 12~hour period.  With very few short baselines and poor \emph{uv}-coverage, resulting \emph{rms} of 0.2~mJy~beam$^{-1}$ is the highest amongst all the analysed observations. Consequently, the data for this observation are of very poor quality and while the results are presented here, we exclude measurements from this image in our determination of expansion rates and age. 
The archival VLA observations were from the 1984 (Project AG0146), 1989 (Project AB0544) and 2008 (project AG0793), see Table~2 for details.

All these observations of \SNR\ were reduced and analysed with the \textsc{miriad} (Sault, Teuben \& Wright 1995) and \textsc{karma} (Gooch 1995) software packages. 

As the resolutions of the produced images  varied (due to the various array configuration, observational periods and therefore resultant \emph{uv}-coverage) the resolutions of all the images were smoothed/convolved to match the image with the lowest resolution ($9\arcsec\times4\arcsec$ at a PA of $0\degr$), see Fig.~4.

Since the $rms$~noise of the resultant images also varied (see Table 3), it was decided not to try and determine the expansion by looking at how the shock front moved between epochs, but by looking at how the radially averaged shell profile peak moved from epoch to epoch.

Using RA=$17^\mathrm{h}48^\mathrm{m}45.4^\mathrm{s}$,  Dec=$-27\degr10\arcmin06\arcsec$ as the centre of \SNR\ we produced normalised shell profiles (Fig.~5), averaged over all angles, for each of the images shown in Fig.~4.\\

\noindent
\parbox{\columnwidth}{
{\bf Table 3.} \SNR\ \emph{rms} noise (1$\sigma$) of 6~cm images shown in Fig.~4. All five images shown in Fig.~4 have matched resolution of 9\arcsec$\times$4\arcsec\ at a PA of $0\degr$.\\
\vskip.25cm
\centerline{
 \begin{tabular}{@{}llc}
 \hline
Telescope &   Date & Image \emph{rms} \\
                 &           & (mJy/beam)\\
 \hline       
VLA   &  1984 & 0.09    \\ 
VLA   &    1989  & 0.09 \\ 
ATCA &  1993 & 0.20 \\ 
VLA   &    2008 & 0.06 \\ 
ATCA &  2009 & 0.06 \\ 
  \hline        
\end{tabular}}} 
\vskip.5cm

From the shell profiles in Fig.~5 we have determined the expansion rate in arc-seconds per year, percentage~expansion per year,  the averaged expansion velocity in km\,s$^{-1}$ (assuming a distance of 8.5~kpc\footnote{Distance estimates to \SNR\ vary between $\sim$7.8~kpc (Nord et al. 2004) and 10~kpc  (Roy \& Pal 2014). We use 8.5~kpc as the most likely distance -- suggested by Reynolds et al. (2008).}), and age in years. These results are summarised in Tables 4 though 7 respectively.

\end{multicols}

\centerline{
\includegraphics[width=130mm,angle=0]{fig4}}
\label{fig3}
\figurecaption{4.}{Matched resolution ($9\arcsec\times4\arcsec$ at a PA of $0\degr$) 6-cm images of Galactic SNR \SNR\ (centred at RA(J2000)=$17^\mathrm{h}48^\mathrm{m}45.4^\mathrm{s}$, Dec(J2000)=$-27\degr10\arcmin06\arcsec$) at multiple epochs. Left to right, top to bottom, 2009 ATCA, 2008 VLA, 1993 ATCA, 1989 VLA \& 1984 VLA. 
}


\centerline{
\includegraphics[width=160mm,angle=0]{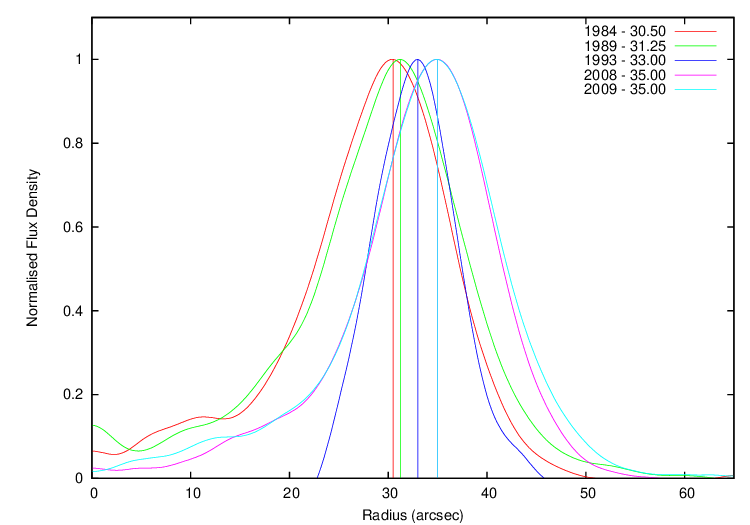}}
\label{fig3}
\figurecaption{5.}{Normalised shell profiles averaged over all angles for each image of \SNR\ in Fig.~4.}

\begin{multicols}{2}

\noindent
\parbox{\columnwidth}{
{\bf Table 4.} Expansion rate of \SNR\ in arc-seconds per year of radially averaged radio-continuum emission peak. \\
\vskip.25cm
\centerline{
 \begin{tabular}{@{}lrrrr}
 \hline
   & 1984 & 1989 & 2008 & 2009\\
 \hline       
  1984 & 0  \\
  1989 & $0.148$ & 0 \\
  2008 & $0.189$ & $0.200$  & 0 \\
  2009 & $0.182$ & $0.191$  & 0 & 0 \\
 \hline        
\end{tabular}}} 
\vskip.5cm

\noindent
\parbox{\columnwidth}{
{\bf Table 5.} \% Expansion of \SNR\ per year of radially averaged radio-continuum emission peak. \\
\vskip.25cm
\centerline{
 \begin{tabular}{@{}lrrrr}
 \hline
   & 1984 & 1989 & 2008 & 2009\\
 \hline       
  1984 & 0  \\
  1989 & $0.484$ & 0 \\
  2008 & $0.620$ & $0.641$  & 0 \\
  2009 & $0.598$ & $0.612$  & 0 & 0 \\
 \hline        
\end{tabular}}} 
\vskip.5cm

noindent
\parbox{\columnwidth}{
{\bf Table 6.} Estimated age of \SNR\ in years. \\ 
\vskip.25cm
\centerline{
 \begin{tabular}{@{}lrrrr}
 \hline
   & 1984 & 1989 &  2008 & 2009\\
 \hline       
  1984 & 0  \\
  1989 & $206$ & 0 \\
  2008 & $161$ & $156$ &  0 \\
  2009 & $167$ & $163$ &  0 & 0 \\
 \hline        
\end{tabular}}} 
\vskip.5cm

\noindent
\parbox{\columnwidth}{
{\bf Table 7.} Speed of radially averaged radio-continuum emission peak in km\,s$^{-1}$ of \SNR. \\
\vskip.25cm
\centerline{
 \begin{tabular}{@{}lrrrr}
 \hline
   & 1984 & 1989 & 2008 & 2009\\
 \hline       
  1984 & 0  \\
  1989 & 5\,950 & 0 \\
  2008 & 7\,620 & 8\,070 & 0 \\
  2009 & 7\,350 & 7\,710 & 0 & 0 \\
 \hline        
\end{tabular}}} 
\vskip.5cm

In Tables 4 and 5 we show the expansion of \SNR\ over the period between 1984 and 2009. The mid range expansion over this period (excluding the 1993 observations) is 0.563\%$\pm$0.078\%~yr$^{-1}$ or 0.174$\pm$0.026~arcsec~yr$^{-1}$. However, we note that between 1984 and 1989, \SNR\ expanded at a somewhat slower rate (0.484\%~yr$^{-1}$) which was also estimated by G\'omez \& Rodr\'iguez (2009) at $0.46\%\pm0.11\%$ based on 20-cm VLA data from the same period. Since 1989, \SNR\ appears to expand at a faster rate of up to 0.641\% per year which is in excellent agreement with the X-ray estimates of Carlton et al. (2011a,b) of 0.642\%$\pm$0.049\% and Green et al. (2008) using VLA observations. This may be an indication that perhaps the shock has``broken through" to a region of lower density and thus accelerated. The mid range expansion rate  derived here,  yields an upper age limit of 181$\pm$25 years (Table~6) assuming  a constant expansion rate since the SN event. This dates the SN event back to the year 1828 ($\pm$25).

Similarly, the speed at which the radially averaged radio-continuum emission peak is moving (assuming a distance to \SNR\ of 8.5~kpc) is somewhat slower in the 1980's ($\sim$6\,000~km~sec$^{-1}$) than later when it speeds up to $\sim$8\,000~km~sec$^{-1}$ (Table~7). The speed of radially averaged radio-continuum emission peak should not be confused with the expansion velocity of the SNR shock front, which has been estimated at $\sim$18\,000~km sec$^{-1}$ by Borkowski et al. (2013). 

As our determination of the radially averaged radio-continuum emission peak speed is dependent on the distance to the SNR we estimate that at the lower end (between 1984 and 1989) the speed varies between 5\,500~km~sec$^{-1}$ (at a distance of 7.8~kpc) and 7\,000~km~sec$^{-1}$ (at a distance of 10~kpc). More recently (between 1989 and 2009), the speed varies between 7\,400~km~sec$^{-1}$ (at a distance of 7.8~kpc) and 9\,500~km~sec$^{-1}$.

\end{multicols}

\centerline{
\includegraphics[width=0.365\columnwidth,angle=-90]{fig6a}
\includegraphics[width=0.365\columnwidth,angle=-90]{fig6b}}
\centerline{
\includegraphics[width=0.365\columnwidth,angle=-90]{fig6c}
\includegraphics[width=0.365\columnwidth,angle=-90]{fig6d}}
\label{fig6}
\figurecaption{6.}{The Spectral Index map of \SNR\ overlaid with 20-cm contours (top left; contours are: 3, 5, 8, 12, 12, 17, 23, 30, 38, 47, 57 $\times\ 7.5\times10^{-4}$ Jy/beam), 13-cm contours (top right; contours are: 3, 5, 8, 12, 17, 23, 30, 38, 47, 57 $\times\ 6.0\times10^{-4}$ Jy/beam), 6-cm contours (bottom left; contours are: 3, 5, 8, 12, 17, 23, 30 $\times\ 5.0\times10^{-4}$ Jy/beam) and 2007 Chandra X-ray contours (bottom right; contours are: 33.57, 67.14, 100.7, 134.3, 167.9, 201.4, 235, 268.6~cts/s).
}

\begin{multicols}{2}

\subsection{3.2 \SNR\ Spectral Energy Distribution}

By matching the resolutions of our 2009 ATCA images at 6-cm and 13-cm to that of the \mbox{20-cm} image ($10\farcs9 \times  5\farcs4$ at PA=--$0\fdg5$), a three point spectral index map was created, allowing for the examination of the spatial spectral variations in the remnant (see Fig.~6). In this map, the colour of each pixel represents the spectral index $\alpha$ across the three observational frequencies.  

From this map we can see that the radio-continuum spectral energy distribution across the NW and SE regions is flatter \mbox{($\alpha\sim-0.5$)} which is also following the contours of the strongest of X-ray emission (so called ``X-ray ears'' of \SNR; see Fig.~6 -- bottom right). This SED flattening in the NW and SE is also confirmed by Farnes (2012) in his VLA observations. The steeper ($\alpha\sim-1$) radio spectra is dominant in the Northern region of \SNR\, corresponding to where the radio emission is the strongest images and therefore indicating synchrotron radio-continuum emission. We also note that somewhat steeper spectral index ($\alpha < -1.0$) is dominating inside part of the SNR while flatter spectra is at the edges.

We estimate that the average spectral index across the entire SNR is $\alpha=-0.72\pm0.26$ which is flatter than LaRosa et al. (2000) ($\alpha=-0.93\pm0.25$), however, their estimates are based on 20/90~cm flux density measurements. This may indicate that the synchrotron emission may even more dominant at higher wavelengths (92~cm). Using Green et al. (2008) flux density estimates at 20 and 6~cm from their 2008 VLA observations we estimate spectral index of $\alpha=-0.62\pm0.08$ which is in good agreement with our ATCA estimates from approximately one year later. This steeper spectral index is expected for younger SNRs (Bell et al. 2011) and further confirms its young age.

\subsection{3.3 Polarisation of \SNR}

Since the ATCA observations recorded Stokes parameters $Q$, $U$ and $V$, in addition to total intensity $I$, we were able to determine the polarisation of \SNR. In our 6~cm image (Fig.~7) we show the regions of polarised emission for \SNR. The electric field vectors follow the shell of the SNR around most of the circumference of the SNR, particularly along its eastern side. 

The maximum fractional polarisation is estimated to be $P=17\pm3\%$ with a mean of $6\%$. No reliable polarisation was detected at 20 or 13~cm. This might indicate significant depolarisation in the remnant, however the polarimetric response of the ATCA is known to be poor at 13-cm. 

   \centerline{
   \includegraphics[width=0.375\textwidth,angle=-90]{fig7}
   }
\label{fig7}
\figurecaption{7.}{6-cm ATCA observations of \SNR. The blue ellipse in the lower-left corner represents the synthesised beam width of $10\farcs940 \times 5\farcs384$ at \mbox{PA=--$0\fdg5$}. The length of the vectors represents the fractional polarised intensity at each pixel position, and their orientations indicate the mean PA of the electric field (averaged over the observing bandwidth, not corrected for any Faraday rotation). The blue line below the beam ellipse represents the length of a polarisation vector of $100\%$. The maximum fractional polarisation is $17\%\pm3\%$ with a mean of $6\%$. Contours at 2.2, 2.8, 4.4, 6.6, 9.4, 13, 17, 21, 26  and 32~mJy~beam$^{-1}$).}

Typically, young type~Ia SNRs exhibit a radially oriented magnetic field (tangentially oriented electric field), which is to be expected from Rayleigh-Taylor instabilities in a decelerating remnant (Gull 1975; Chevalier 1976). This is consistent with similarly young Galactic SNRs, as well as in the LMC (e.g. Table~3 in Bozzetto et al. (2014)). As we have plotted electric field vectors we would expect that they would be tangential to the circumference of the remnant. We can see in Fig.~7, the orientation of the electric field vectors roughly follow this arrangement, however it can be seen that this pattern is not strictly followed around the entire remnant. Given the location of \SNR, towards the Galactic centre, this is most likely due to Faraday depolarisation.

Farnes (2012) also detected the presence of polarised emission. Our ATCA polarimetric results also suggest that the above observed variation is most consistent with an ambient B field perpendicular to the axis of bilateral symmetry indicated by Farnes (2012). Moreover, Farnes (2012) argues that increased ordering of the B field in the NW as the strong Faraday depolarisation must also be present. 

Farnes (2012), also argues that an intrinsically radially-oriented field could be provided by a systematic gradient in Rotation Measure (RM) of 140~rad~m$^{-2}$ from N to S and can also explain the depolarisation that we observe in our ATCA images (see Sect.~3.3). 

\subsection{3.4 Magnetic Field of \SNR}

We used the new equipartition formulae derived by Arbutina et al. (2012, 2013) based on the diffusive shock acceleration (DSA) theory of Bell (1978) to estimate a magnetic field strength. These formulae are particularly relevant to magnetic field estimation in SNRs, and yields magnetic field strengths between those given by the classical equipartition (Pacholczyk 1970) and revised equipartition (Beck and Krause, 2005) methods. We estimate the magnetic field strength of \SNR\ to be B $\approx 273~\mu\mbox{G}$ and the minimum total energy of the synchrotron radiation to be \mbox{E$_{\textrm{min}} \approx 1.8\times10^{48}~\mbox{ergs}$} (see Arbutina et al. (2012, 2013)); and corresponding online calculator\footnote{http://poincare.matf.bg.ac.rs/$\sim$arbo/eqp/}). For this estimate, we used a spectral index value of \mbox{$\alpha = -0.72$}, integrated flux density S$_{1.425}$=0.935~Jy at $\nu$=1.425~GHz (Green et al. 2008), distance D=8.5~kpc, SNR radius of r=46\arcsec\ and filling factor of 0.33. However, if we additionally assume shock velocity of 18\,000~km~s$^{-1}$ (as suggested by Borkowski et al. (2013)) than the magnetic field strength of \SNR\ becomes somewhat lower (B $\approx 180~\mu\mbox{G}$ and the minimum total energy of the synchrotron radiation is \mbox{E$_{\textrm{min}} \approx 7.6\times10^{47}~\mbox{ergs}$}. These estimates are very similar to Arbutina et al. (2012) estimates of \mbox{B $\approx~228~\mu\mbox{G}$} and \mbox{E$_{\textrm{min}} \approx 9.3\times10^{47}~\mbox{ergs}$}. 

A large magnetic field strength such as this ($273~\mu\mbox{G}$), is expected for a young SNR (Bell~2004). Indeed, this makes \SNR, a remnant with one of the the strongest estimated magnetic field strengths known to date. For example, other young Galactic SNRs (Beck \& Krause~2005; Arbutinal et al. 2012 (see their Table~1)) such as Cas~A (\mbox{B$\approx~1250~\mu\mbox{G}$}), Kepler (\mbox{B$\approx~414~\mu\mbox{G}$}), G349.7+0.2 (\mbox{B$\approx~523~\mu\mbox{G}$}) and Tycho (\mbox{B$\approx~285~\mu\mbox{G}$}) are known remnants with the strongest magnetic fields. Also, in a Large Magellanic Cloud (LMC) SNR, J0509-6731 (also remnant from a Type~Ia SN explosion) at about 400~yrs age has magnetic field strength of $168~\mu\mbox{G}$ (Bozzetto et. al. 2014) while the magnetic field of LMC~SNR~J0519--6902  is $171~\mu\mbox{G}$ (Bozzetto et al. 2012). The Small Magellanic Cloud SNR HFPK~443 (Crawford et al. 2014, in press) has a field strength of $90~\mu\mbox{G}$ with numerous other older remnants falling below these values. It is most likely that \SNR\ is going through an evolutionary stage where the magnetic field has been amplified (added to simple compression by the shocks), which may explain such a high magnetic field value (Telezhinsky et al. 2012). The amplification of magnetic field is a process driven by the very fast shocks of young SNRs. Because of strong amplification of magnetic field, a spectral index of $\alpha=-0.72$ and the location in the surface brightness-diameter diagram this SNR is of younger age, in free expansion stage and expanding in a low density environment. 

In Fig.~8 we show a surface brightness--diameter ($\Sigma-D$) diagram at 1~GHz with theoretically-derived evolutionary tracks (Berezhko \& V\"olk 2004) superposed. \SNR\ lies at $(D,\Sigma)$ = (3.8~pc, $7.5\times 10^{-20}$~W m$^{-2}$~Hz$^{-1}$~Sr$^{-1}$) on the diagram. Its position on the diagram tentatively suggests that it is in the mid-to-late free expansion phase of evolution ---  expanding into a low density environment  $(n_\mathrm{H}\lesssim 0.3 \mathrm{cm}^{-3})$.

   \centerline{
   \includegraphics[width=0.55\textwidth]{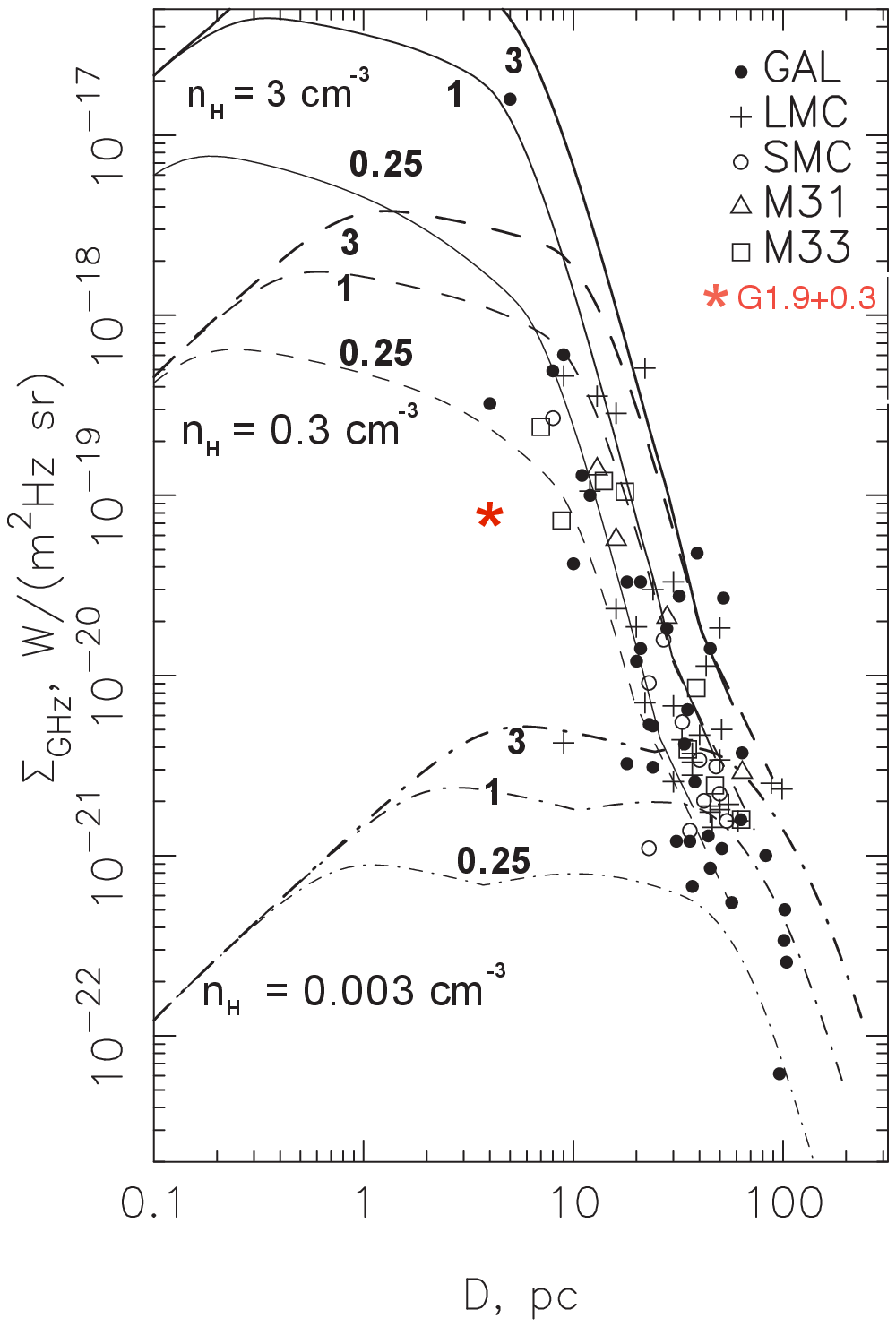}
   }
\label{fig8}
\figurecaption{8.}{1 GHz Surface brightness-to-diameter diagram from  Berezhko \& V\"olk (2004), with \SNR\ added. The evolutionary tracks are for ISM densities of N$_\mathrm{H}$= 3, 0.3 and 0.003~cm$^{-3}$ and explosion energies of E$_{\mathrm{SN}}$ = 0.25, 1 and 3$\times10^{51}$~erg.
}

\section{4. CONCLUSIONS}

Here, we have presented new 6-cm, 13-cm and 20-cm ATCA observation of Galactic SNR \SNR\ made in 2009. Using the new 6-cm data and archival 6-cm data, we observe that there are indications that the expansion of \SNR\ accelerated after 1989. Our results are in broad agreement with the estimates of expansion made by Reynolds et al. (2008),  Green et al. (2008), G\'omez \& Rodr\'iguez (2009) and Carlton et al. (2011a,b). We find that at $\sim181$~yrs, \SNR\ is indeed very young and most likely the youngest SNR in the Milky Way. This very young age is also reinforced by magnetic field arrangement and strength we estimate.    \\

\noindent We make the following findings:\\

\item - Expansion rate of 0.484\% per year between 1984 and 1989;
\item - Expansion rate of 0.641\% per year between 1989 and 2009;
\item - Expansion rate  of 0.563\%$\pm$0.078\% per year between 1984 and 2009;
\item - Age of \mbox{181~yrs$\pm$25~yrs};
\item - Average  spectral index between 20-cm and 6-cm, across the entire SNR is $\alpha$=--0.72$\pm$0.26;
\item - 6\% fractionally polarised radio emission with a peak of 17\%$\pm$3\%;
\item - Magnetic field strength B$\approx$273~$\mu\mbox{G}$;
\\

\acknowledgements{We used the {\sc karma} and {\sc miriad} software package developed by the ATNF. The Australia Telescope Compact Array is part of the Australia Telescope which is funded by the Commonwealth of Australia for operation as a National Facility managed by CSIRO. The Karl G. Jansky Very Large Array (VLA). The National Radio Astronomy Observatory is a facility of the National Science Foundation operated under cooperative agreement by Associated Universities, Inc. 
}

\references

Abramowski, A. and H. E. S. S. Collaboration et al.: 2014, \journal{Mon. Not. R. Astron. Soc.}, \vol{441}, 790.

Arbutina B., Uro\v sevi\'c D., Andjeli\'c M. M., Pavlovi\'c, M. Z., Vukoti\'c B.: 2012, \journal{Astrophys. J.}, \vol{746}, 79.

Arbutina B., Uro\v sevi\'c D., Vu\v ceti\'c M. M., Pavlovi\'c, M. Z., Vukoti\'c B.: 2013, \journal{Astrophys. J.}, \vol{777}, 31.

Beck, R., and Krause, M.: 2005, \journal{Astron. Nachr.}, \vol{326}, 414.

Bell, A. R.: 1978, \journal{Mon. Not. R. Astron. Soc.}, \vol{182}, 443.

Bell, A. R.: 2004, \journal{Mon. Not. R. Astron. Soc.}, \vol{353}, 550.

Bell, A. R., Schure K. M., Reville B.: 2011, \journal{Mon. Not. R. Astron. Soc.}, \vol{418}, 1208.

Berezhko, E. G. and V\"olk, H. J.: 2004, \journal{Astron. Astrophys.}, \vol{427}, 525.

Borkowski, K. J., Reynolds, S. P., Hwang, U., Green, D. A., Petre, R., Krishnamurthy, K. and  Willett, R.: 2013, \journal{Astrophys. J.}, \vol{771}, L9.

Bozzetto, L. M., Filipovi\'c, M. D., Uro\v sevi\'c, D. and Crawford, E. J.: 2012, \journal{Serb. Astron. J.}, \vol{185}, 25.

Bozzetto, L. M., Filipovi\'c, M. D., Uro\v sevi\'c, D., Kothes, R. and Crawford, E. J.: 2014, \journal{Mon. Not. R. Astron. Soc.}, \vol{440}, 3220.

Cappellaro, E.:  2003, in "Supernovae and Gamma-Ray Bursters", ed. Weiler K., Lecture Notes in Physics, Berlin Springer Verlag, Vol. \vol{598}, 37.

Carlton, A. K., Borkowski, K. J., Reynolds, S. P., Willett, R., Krishnamurthy, K., Green, D. A. and  Petre, R.: 2011a, in American Astronomical Society Meeting Abstracts 217. p. 256.12.

Carlton, A. K., Borkowski, K. J., Reynolds, S. P., Hwang, U., Petre, R., Green, D. A., Krishnamurthy, K. and Willett, R.: 2011b, \journal{Astrophys. J.}, \vol{737}, L22.

Chevalier, R. A.: 1976, \journal{Astrophys. J.}, \vol{207}, 872.

Farnes, J., 2012, PhD thesis, University of Cambridge.

Gooch, R. E.: 1995, in Astronomical Data Analysis Software and Systems IV, ASP Conf. Series ed. Shaw, R. A., Payne, H. E. and Hayes, J. J. E., \vol{77}, 144. 

Gray, A. D.: 1994, \journal{Mon. Not. R. Astron. Soc.}, \vol{270}, 835.

Green, D. A. and  Gull, S. F.: 1984, \journal{Nature}, \vol{312}, 527.

Green, D. A.: 2004, \journal{Bull. Astr. Soc. India}, \vol{32}, 335.

Green, D. A., Reynolds, S. P., Borkowski, K. J., Hwang, U., Harrus, I. and Petre, R.: 2008, \journal{Mon. Not. R. Astron. Soc.}, \vol{387}, L54.

G\'omez, Y. and  Rodr\'iguez, L. F.: 2009, \journal{Rev. Mex. Astron. Astrofis.}, \vol{45}, 91.

Gull, S. F: 1975, \journal{Mon. Not. R. Astron. Soc.}, \vol{171}, 263.

LaRosa, T. N., Kassim, N. E., Lazio, T. J. W. and Hyman, S. D.: 2000, \journal{Astron. J.}, \vol{119}, 207.

Murphy, T., Gaensler, B. M., Chatterjee, S.: 2008, \journal{Mon. Not. R. Astron. Soc.}, \vol{389}, L23.

Nord, M. E., Lazio, T. J. W., Kassim, N. E., Hyman, S. D., LaRosa, T. N., Brogan, C. L. and  Duric, N.: 2004, \journal{Astron. J.}, \vol{128}, 1646.

Pacholczyk, A. G.: 1970, in "Radio astrophysics. Nonthermal processes in galactic and extragalactic sources", ed. A. G. Pacholczyk (San Francisco: Freeman).

Reynolds, S. P., Borkowski, K. J., Green, D. A., Hwang, U., Harrus, I. and  Petre, R.: 2008, \journal{Astrophys. J.}, \vol{680}, L41.

Reynolds, S. P., Borkowski, K. J., Green, D. A., Hwang, U., Harrus, I. and  Petre, R.: 2009, \journal{Astrophys. J.}, \vol{695}, L149.

Roy, S. and  Pal, S.: 2014, in "Supernova environmental impacts", eds. Ray A. and McCray R. A., IAU Symposium Vol. \vol{296}, 197.

Sault, B. and  Killeen, N.: 2008, Miriad Users Guide. Australia Telescope National Facility.

Sault, R. J., Teuben, P. J. and Wright, M. C. H.: 1995, in "Astronomical Data Analysis Software and Systems IV ASP Conference Series", eds. Shaw, R. A., Payne, H. E. and Hayes, J. J. E., \vol{77}, 433.

Telezhinsky, I., Dwarkadas, V. V., and Pohl, M.: 2012, \journal{Astropart. Phys.}, \vol{35}, 300.

van den Bergh, S. and  Tammann, G. A.: 1991, \journal{Annu. Rev. Astron. Astrophys.}, \vol{29}, 363.

\endreferences

\end{multicols}

\vfill\eject

{\ }



\naslov{RADIO KONTINUUM EMISIJA MLADOG GALAKTIQKOG OSTATKA SUPERNOVE \textrm{\SNR}}


\authors{A.~Y.~De~Horta$^{1}$, M.~D.~Filipovi\'c$^{1}$, E.~J.~Crawford$^{1}$, F.~H.~Stootman$^{1}$, T.~G.~Pannuti$^{2}$,}
\authors{L.~M.~Bozzetto$^{1}$, J.~D.~Collier$^{1}$, E.~R.~Sommer$^{1}$ and A.~R.~Kosakowski$^{2}$}
\vskip3mm


\address{$^{1}$University of Western Sydney, Locked Bag 1797, Penrith South, DC, NSW 1797, Australia }

\Email{a.dehorta}{uws.edu.au, m.filipovic@uws.edu.au, e.crawford@uws.edu.au}

\address{$^{2}$Space Science Center, Department of Earth and Space Sciences, Morehead State University, 235 Martindale Drive, Morehead, KY 40351 USA}

\Email{t.pannuti}{moreheadstate.edu}

\vskip.7cm


\centerline{UDK \udc}


\centerline{\rit Uredjivaqki prilog}

\vskip.7cm

\begin{multicols}{2}
{


{\rrm U ovoj studiji predstav{lj}amo nova radio-kontinuum posmatranja na 6~cm najmla{\dj}eg ostatka supernovih (OS) u naxoj Galaksiji. Ostatak \textrm{\SNR} je \mbox{$\sim$181$\pm$25} godina star i godi{\ss}{\nj}a stopa rasta mu je 0.563\%$\pm$0.078\% u periodu izme{\dj}u 1984 i 1989. Primetili smo da se izme{\dj}u 1984 i 1989, ostatak {\ss}irio ne{\ss}to sporije (0.484\% godi{\ss}{\nj}e) nego posle 1989-te kada je ekspanzija dostigla nivo od 0.641\% godi{\ss}{\nj}e. \textrm{\SNR} ima tipiqan radio-spektar za mlade ostatke sa \mbox{$\alpha=$--0.72$\pm$0.26.} \textrm{\SNR} emituje proseqno 6\% polarizovanog zraqenja na talsnoj du{\zz}ini od 6~cm, sa maksimalnim intenzitetom 17\%$\pm$3\%. Ova polarizaciona emisija tako{\dj}e tesno prati rentgensko zraq{\nj}e. Proce{nj}eno je priliqno jako magnetno po{lj}e u vrednosti od 273~$\mu${\rm G} {\ss}to predstav{\lj}a jedno od najsna{\zz}nijih magnetnih po{\lj}a u do sada posmatranim OS.}

}
\end{multicols}

\end{document}